\input phyzzx
\hoffset=0.375in
\overfullrule=0pt

\def\dol{{d_{\rm ol}}}
\def\dls{{d_{\rm ls}}}
\def\dos{{d_{\rm os}}}
\def\max{{\rm max}}

\def\kpc{{\rm kpc}}
\def\pc{{\rm pc}}

\twelvepoint
\font\bigfont=cmr17
\centerline{\bigfont LMC Microlenses: Dark or Luminous?}
\bigskip
\centerline{{\bf Andrew Gould}\footnote{1}{Alfred P.\ Sloan Foundation Fellow}}
\smallskip
\centerline{Dept of Astronomy, Ohio State University, Columbus, OH 43210}
\smallskip
\centerline{gould@astronomy.ohio-state.edu}
\bigskip
\centerline{\bf Abstract}
\singlespace 

Zhao has proposed that the microlensing events observed toward the Large
Magellanic Cloud (LMC) could be due to faint stars in a dwarf galaxy or 
tidal debris lying along the line of sight to the LMC.  Zaritsky \& Lin claim
to have detected such a structure which, they believe, could account for
most of the observed microlensing optical depth.  Here I show that 
a large-area surface-brightness map made by
de Vaucouleurs constrains any such structure to one of four
possibilities.  Either 1) it does not account for a significant fraction
of the observed microlensing, 2) it covers the inner $\sim 3^\circ$ of
the LMC but does not extend beyond $\sim 5^\circ$ from the LMC center,
3) it is smooth on scales of $\sim 15^\circ$ in both transverse directions  or
4) it has a stellar mass-to-light ratio which exceeds by a factor $\gsim 10$
that of known stellar populations.  The second and third possibilities would
not be expected to apply to tidal debris.  The last merely rephrases the 
dark-matter problem in a new form.

\bigskip
Subject Headings: dark matter -- Galaxy: halo -- gravitational lensing 
-- Magellanic Clouds

\endpage
\chapter{Introduction}

	The MACHO (Alcock et al.\ 1997a) and EROS (Aubourg et al.\ 1993)
collaborations have detected a total of 16 candidate microlensing events
toward the Large Magellanic Cloud (LMC).  While a few of these may be
variable stars, the great majority are likely to be genuine microlensing.
MACHO has estimated an optical depth of $\tau=2.9^{+1.4}_{-0.9}\times 10^{-7}$
based on a subset of these detections, a significant fraction of the
$\tau\sim 4.7\times 10^{-7}$ expected if the dark halo of the Milky Way were
composed entirely of massive compact halo objects (MACHOs).  Sahu (1994)
suggested that a large fraction of the events could be due to lensing
by stars within the LMC itself, particularly in the LMC bar, and indeed one 
event appears to be due to a binary in the LMC 
(Alcock et al.\ 1997a).  However, general dynamical arguments constrain the
self-lensing of the LMC disk to $\tau_{\rm self}\lsim 1\times 10^{-8}$
(Gould 1995). Moreover, as observations have continued it is becoming 
apparent that the events do not occur preferentially in the bar which is 
what one would expect if they were due primarily to LMC stars.  Some events
are also expected from stars in the Milky Way disk.  However, based on star 
counts, Gould, Bahcall, \& Flynn (1997) estimate 
$\tau_{\rm MW}\sim 8\times 10^{-9}$.  This is about 35 times smaller than the
observed value, although Gould et al.\ (1997) argue that one of the observed
events may well be due to a disk M dwarf.  In brief, the majority of these
events do not seem to be due to known stellar populations.  This is an
important and puzzling result because the estimate of the typical mass of
the lenses (derived from the observed timescales of the events and dynamical
models of the Galactic halo) is $\sim 0.4\,M_\odot$.  If these objects were
composed of hydrogen they would burn, and the population would easily be
detected (assuming it were distributed thoughout the Galaxy).  
If they are some new exotic object, it is most curious that they
have the mass of normal stars.

	This puzzle led Zhao (1997) to suggest that the events may be
due to a dwarf galaxy or tidal debris from a disrupted galaxy along the line
of sight to the LMC.  This would appear to explain in a natural way why the
inferred masses are similar to those of stars: the lenses are stars.  More
recently, Zaritsky \& Lin (1997)
claim to have detected such a foreground structure (but see Alcock et al.\ 
1997b).
They observed a field $\sim 2^\circ$ northwest of the LMC bar and found that
$\sim 5\%$ of the clump giants are brighter than the mean by 0.9 mag.  They
interpret this as evidence for a stellar population 0.9 mag in the foreground.
They estimate a stellar surface mass density of $\Sigma\sim 16\,M_\odot\,\rm 
pc^{-2}$ which would account for a large fraction of the microlensing events.
This conclusion is very appealing in that it would eliminate the need for
exotic objects.

	However, the problem of the ``dark matter'' apparently being detected
in microlensing experiments is not
diminished by positing that this mass lies in previously unrecognized 
structures.  To be a natural solution to the problem, these structures must
also have normal mass-to-light ratios.  If newly found structures such as the
one claimed by Zaritsky \& Lin (1997) were in fact responsible for the
microlensing events, but they had anomalously high mass-to-light ratios,
the ``dark matter'' mystery would simply take on a new form.  I therefore 
investigate what limits can be placed using existing data on the 
mass-to-light ratio of previously unrecognized structures 

\chapter{Mass and Light}

	If a structure, particularly tidal debris, lay 10 or 20 kpc in front
of the LMC, one would not expect its angular extent to be perfectly
coincident with that of the LMC.  In general it should, like the Magellanic
Stream, extend well beyond the LMC.  The surface brightness
of any structure extending beyond $\sim 5^\circ$ from the LMC center
is constrained by the LMC surface-brightness map of 
de Vaucouleurs (1957) 
to be fainter than $R\gsim 25\rm 
\,mag\,arcsec^{-2}$, the last isophote of
the map.  Assuming $V-R\sim 0.5$, this corresponds to
$V\gsim 25.5\rm\,mag\,arcsec^{-2}$, or a surface brightness of
$$S_\max \sim 2.2\, L_\odot \,\rm pc^{-2},\eqn\sblim$$
where I have used the identity
$$V = 26.4\,{\rm mag}\,{\rm arcsec}^{-2}\Leftrightarrow
S = 1 L_\odot\,\pc^{-2}.\eqn\ident$$

	On the other hand, to account for a fraction $f$ of the observed
microlensing optical depth $\tau=2.9\times 10^{-7}$ by a stellar structure
at a distance $\dol$ requires a surface density $\Sigma$ 
$$\Sigma = {f\tau c^2\over 4 \pi G D} = 
47 f \biggl({D\over 10\,\kpc}\biggr)^{-1}
M_\odot\,\pc^{-2},\eqn\sigest$$
where $D\equiv \dol\dls/\dos$ and $\dol$, $\dls$, and $\dos$ are the distances
between the observer, the lensing structure, and the LMC sources.
Hence, the stellar mass-to-light ratio of the structure must exceed
$${M\over L} > 22 f \biggl({M\over L}\biggr)_\odot= 12\,f\,
\biggl({M\over L}\biggr)_{\rm MW},
\eqn\moverl$$
where I have normalized the mass-to-light ratio to that 
observed for stars in the disk of the
Milky Way.  This is obtained by dividing the observed
stellar column density of the local
disk $\Sigma_{\rm MW}=27\,M_\odot\,\pc^{-2}$ (Gould, Bahcall, \& Flynn 1996, 
1997) by the observed surface brightness of the local
disk, $S_{\rm MW}= 15\,L_\odot\,\pc^{-2}$ (Binney \& Tremaine 1987), that is
$(M/L)_{\rm MW} =\Sigma_{\rm MW} /S_{\rm MW} =
1.8(M/L)_{\odot}$.

\chapter{Discussion}

	Equation \moverl\ implies one of four conclusions.  Either:

1) The structures along the line of sight toward the LMC have a normal stellar
mass-to-light ratio but {\it in total} account for only
a fraction $f\lsim 1/12$ of the observed microlensing events.

2) The argument of \S\ 2 fails because the intervening structures happen to
be contained within $5^\circ$ of the LMC center (although they must extend 
over at least the
inner $\sim 3^\circ$ to account for the observed events).  This possibility 
could only apply to 
a self-gravitating structure and not to an intrinsically extended one such as 
tidal debris.

3) The argument of \S\ 2 fails because the intervening material is smooth
on scales of $\sim 15^\circ$, the size of de Vaucouleurs' (1957) map.
Before constructing the LMC isophotes, de Vaucouleurs had to remove a smooth
foreground component which has a mean surface brightness of 
$R\sim 21.2\,\rm mag\, arcsec^{-2}$ with a gradient of $\sim 15\%$ across the
field.  The mean surface brightness is primarily due to the sky and 
de Vaucouleurs (1957) assumed that the gradient is due to Galactic foreground.
(The direction of the gradient is toward the Galactic plane.)\ \  However,
if there were any other foreground structures that were smooth on the scales of
the map, these would have been removed also.

4) The structures along the line of sight have a mass-to-light ratio an order 
of magnitude higher than
the local disk.  They could then account for the observed microlensing events,
but would be composed primarily of compact dark objects.

	How do these conclusions square with the claims of Zaritsky \&
Lin (1997) to have detected a structure with $\Sigma=16\,M_\odot\,\pc^{-2}$?
First note that at the center of the field, the LMC has a surface brightness
$R\sim 22.4\,\rm mag\,arcsec^{-2}$ (de Vaucouleurs 1957).  
The foreground structure has $\gsim 20$
times fewer stars, but these are 0.9 mag brighter, implying a surface
brightness $R\sim 24.7\,\rm mag\,arcsec^{-2}$ which is close to
the limit derived in \S\ 2.  

	There are, however, several problems with the mass estimate, which
taken at face value implies $M/L = 6\,(M/L)_\odot$.  First,
in deprojecting the LMC disk (assumed to be inclined at $i\sim 33^\circ$)
Zaritsky \& Lin (1997) 
used $\csc i$ rather than $\sec i$.  If one carries through their
calculation but making this one correction, one finds an LMC density
of $\Sigma=103\,M_\odot\,\pc^{-2}$ rather than 159.  This still implies a
stellar mass-to-light ratio $M/L\sim 4(M/L)_\odot$ which is about twice that of
the Milky Way disk.
Second, they made their calculation
assuming that all the dynamical mass is in stars.  That is, they assumed that
the LMC does not contain any dark matter or gas at the location of their field.
This assumption is sometimes regarded as
plausible for the inner parts of galaxies and, as noted above, would imply
a stellar mass-to-light ratio only about twice that of the Milky Way
disk.  However, given the flat LMC rotation curve, it also
implies that the surface mass density falls inversely with radius.  If
in fact there were no dark matter, this would mean that the light density
should fall at the same rate (assuming a constant mass-to-light ratio).  
Actually according to the map of 
de Vaucouleurs (1957), it falls much faster than this which
indicates substantial quantities of dark matter.  Thus, $M/L<4\,(M/L)_\odot$
and there is therefore no evidence
that the mass-to-light ratio of the LMC stellar disk differs substantially
from that of the Milky Way.  Finally, even if the stellar mass-to-light ratio
were 6 as Zaritsky \& Lin (1997) derived, this would explain less than
half of the optical depth.  In sum, the apparent 10-fold discrepancy between
the estimates of Zaritsky \& Lin (1997) and this paper proves to be a
combination of several factor $\sim 2$ effects.

{\bf Acknowledgements}:  I would like to thank Nathalie Palanque-Delabrouille
who drew my attention to the work of de Vaucouleurs and Piotr
Popowski and Gerald Newsom for careful readings of the manuscript.
This work was supported in part by grant AST 94-20746 from the NSF and in 
part by grant NAG5-3111 from NASA.

\bigskip
\Ref\alc{Alcock et al.\ 1997a, ApJ, 486, 697}
\Ref\alc{Alcock et al.\ 1997b, ApJ, submitted (astro-ph 9707310)}
\Ref\Aubourg{Aubourg, E., et al.\ 1993, Nature, 365, 623}
\Ref\BT{Binney, J.\ \& Tremaine, S.\ 1987, Galactic Dynamics, (Princeton:
Princeton Univ.\ Press)}
\Ref\dVF{de Vaucouleurs, G. 1957, AJ, 62, 69}
\Ref\gthree{Gould, A.\ 1995, ApJ, 441, 77}
\Ref\gthree{Gould, A., Bahcall, J.\ N., \& Flynn, C.\ 1996, ApJ, 465, 759}
\Ref\gthree{Gould, A., Bahcall, J.\ N., \& Flynn, C.\ 1997, ApJ, 482, 913}
\Ref\zarit{Zaritsky, D., \& Lin, D.\ N.\ C. 1997, AJ, in press}
\Ref\zhao{Zhao, H.\ 1997, MNRAS, submitted (astro-ph 9703097)}
\refout
\endpage
\end